\documentclass[twocolumn,showpacs]{revtex4}

\usepackage{graphicx}
\usepackage{dcolumn}
\usepackage{bm}
\usepackage{amsmath}

\begin{document}
\title{The urban economy as a scale-free network}
\author{Claes Andersson}
\affiliation{Physical Resource Theory, Chalmers University of
Technology, 412 96 G\"oteborg,  Sweden, claes@santafe.edu}
\author{Alexander Hellervik}
\affiliation{Physical Resource Theory, Chalmers University of
Technology, 412 96 G\"oteborg,  Sweden, f98alhe@dd.chalmers.se}
\author{Kristian Lindgren}
\affiliation{Physical Resource Theory, Chalmers University of
Technology, 412 96 G\"oteborg,  Sweden, frtkl@fy.chalmers.se}
\author{Anders Hagson}
\affiliation{City and Mobility, Chalmers University of Technology,
412 96 G\"oteborg,  Sweden, hagson@arch.chalmers.se}
\author{Jonas Tornberg}
\affiliation{City and Mobility, Chalmers University of Technology,
412 96 G\"oteborg,  Sweden, jonas@arch.chalmers.se}

\date{\today}

\begin{abstract}
We present empirical evidence that land values are scale-free and
introduce a network model that reproduces the observations. The
network approach to urban modelling is based on the assumption
that the market dynamics that generates land values can be
represented as a growing scale-free network. Our results suggest
that the network properties of trade between specialized
activities causes land values, and likely also other observables
such as population, to be power law distributed. In addition to
being an attractive avenue for further analytical inquiry, the
network representation is also applicable to empirical data and is
thereby attractive for predictive modelling.
\end{abstract}

\maketitle

\section{\label{i}Introduction}
The Zipf rank-size law for city sizes is one of the most widely
known power laws in science\cite{Zipf_1949}. It is also but one
out of many similar power laws from systems in biology, economy
and society. We continue this research by presenting empirical
evidence that land values are scale-free. The data we use is based
on a database delivered by Sweden Statistics that covers
estimations of the market value of all land in Sweden (2.9 million
data points).

Although power laws are common they are not easily reconstructed
from realistic underlaying dynamics. Their ubiquity suggests that
they could be caused by some general systemic property common to a
range of systems. Recent research suggests that many empirically
observed power laws may be due to fundamental properties of these
systems viewed as networks of interacting
nodes\cite{Barabasi_Albert_1999,Albert_Barabasi_2002a,Dorogovtsev_Mendes_2002a}.
We investigate the mechanisms causing land values to follow these
statistics and present a network model that reproduces the
empirical results. The model is based on basic definitions of city
formation in urban economics theory\cite{OSullivan_2002}.

According to urban economics theory, the formation of modern
cities is primarily caused by the advantages of trade between
specialized producers. The exchange of goods and services between
localized and largely immobile activities in trade economies makes
a network representation natural: the nodes are units of land and
the edges represent the exchange of goods and services between
them.

It has been shown that the node degrees of a certain class of
growing networks are power law
distributed\cite{Barabasi_Albert_1999,Dorogovtsev_Mendes_Samukhin_2000}.
This class of networks is important because their growth
mechanisms can be mapped to the microscopic dynamics of several
real-world systems. We demonstrate how trade in an urban system
can be represented as a scale free network and that, as a
consequence, land values can be expected to follow the same
distribution. We also verify that the model retains these
properties when spatial constraints are taken into account. We do
this by using a spatial network model to reproduce the empirically
observed distribution of land values.

The network approach solves a fundamental issue in the problem of
modelling urban systems by representing the system at the level of
the underlaying market structure. This allows us to produce prices
in units of currency rather than undefined and subjective fitness
measures. It thereby opens up doors for several extensions of the
scope of the model and provides a natural interface for
integrating it with other models.

When we refer to an urban system we do not necessarily refer to
individual cities but rather to systems of specialized trading
activities. To clarify further, our use of the term
\emph{activity} refers to trade gains in units of \emph{currency
per unit area and unit time}. Activities can be resolved to any
resolution down to individual transactions.

\section{\label{ueaan}The urban economy as a network}
In section \ref{tm_formulation} we define a non-spatial model of
urban economic growth and in section \ref{tm_tsm} we extend it to
a spatial model where growth is mapped to a 2D surface. In section
\ref{tm_uc} we motivate the model ontology and the basic
assumptions on which we have based the model. In particular we
discuss the connection between node degree and land value and how
the urban system meets the criteria for being a scale-free
network.

\subsection{\label{tm_formulation}Formulation of the non-spatial model}
\begin{figure}
\includegraphics[width=8.5cm]{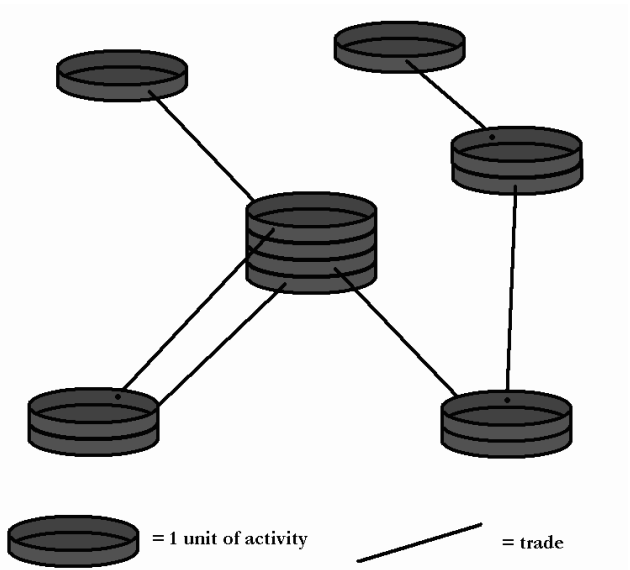}
\caption{\label{networkfig} \em The nodes are non-overlapping
areas of land. Areas with no activity are not shown in the
figure.}
\end{figure}

A geographic area on which an urban economic system can grow is
represented by an enumerated set of nodes, $\{1,2,\ldots, N\}$
corresponding to non-overlapping land areas. Trade of goods and
service between activities in the nodes is represented by
undirected edges. Since activities within the same node can trade
with each other, an edge can connect a node with itself. The
amount of activity of a site $x_i$ is defined as the degree of the
corresponding node. See figure \ref{networkfig}.

The network is initialized by connecting $n_0$ nodes so that each
has a degree of $x_0$. The $N-n_0$ undeveloped nodes have no trade
interactions and thus no activity.

At each time step we update the network as follows:

\begin{enumerate}
\item With probability $q_1$ we add $m$ edges between sites that
already are developed: the first edge end-point is selected
uniformly among developed nodes. The probability of a node $i$ to
be selected is
\begin{equation}
\Pi^u_i=\delta^{(D)}_i \frac{1}{n^{(D)}_t} \label{eq:UNI},
\end{equation}
where $n^{(D)}_t$ is the number of developed nodes at time $t$ and
$\delta^{(D)}_i=1$ if node $i$ is developed and $\delta^{(D)}_i=0$
otherwise.

The second end-point is selected preferentially among developed
nodes. Preferential selection corresponds to the uniform selection
of an activity in the system and the subsequent location of its
node. It was defined by Barabasi et al \cite{Barabasi_Albert_1999}
as
\begin{equation}
\Pi^p_i=\frac{x_i}{\sum_jx_j}, \label{eq:BA}
\end{equation}
where $\Pi^p_i$ is the probability of node $i$ to be attached to a
new edge and $x_i$ is the degree of node $i$.

\item With probability $q_2$ we add edges between $m$ pairs of
nodes that are both selected preferentially according to Eq.
(\ref{eq:BA}).

\item With probability $q_3$ we add $m$ units of initial activity
on land that is previously undeveloped: the first end-point is
selected without bias similarly to Eq. (\ref{eq:UNI}) but among
undeveloped nodes. However, since nodes have no properties beside
their degree in the non-spatial model, any undeveloped node can be
added. The second end-point is selected preferentially according
to Eq. (\ref{eq:BA}).

\end{enumerate}

We will refer to the growth classes as type-1, type-2 and type-3
growth and to their relative rates as $q_1$, $q_2$ and $q_3$ with
$q_1+q_2+q_3=1$ throughout the paper.

\subsubsection{\label{tm_formulation_2}A continuum formulation}

Instead of assuming any particular set of activities and
interactions we can use a continuum formulation where we consider
each edge end-point to be an average of a large set of urban
activities. It follows that an average activity must be assumed to
interact equally much with all other average activities. Rather
than counting explicit interactions to determine the activity
level, we study the evolution of the expected node degrees. The
time evolution of activity on a developed site $i$ follows the
equation
\begin{equation}
x_i(t+1)=x_i(t)+q_1\frac{m}{n^{(D)}_t}+m(q_1+2q_2+q_3)\frac{x_i(t)}{\sum_jx_j(t)}\label{eq:meanBA},
\end{equation}
which is solved by the continuous-time method introduced by
Barab\'asi et al \cite{Albert_Barabasi_Jeong_1999}. After
sufficiently long time the degree distribution approaches the form
\begin{equation}
\label{eq:power} P[x_i=x]\sim(x+A)^{-\gamma},
\end{equation}
with
\begin{equation}
A=\frac{2mq_1}{q_3(1+q_2)}
\end{equation}
and
\begin{equation}
\gamma=1+\frac{2}{1+q_2}\label{eq:exponent}.
\end{equation}

According to Eq. (\ref{eq:power}) the node degree distribution
will be power law distributed for the non-spatial model.

\subsection{\label{tm_tsm}Formulation of the spatial model}

The non-spatial model does not include any information about
inhomogeneous relations in the network. An important departure
from such a simple model that can be readily observed in real
systems is of course the fact that different pairs of sites have
different distances between them. Because of transportation cost
optimization this has the potential to affect the edge
distribution of the network model.

We will now extend the basic model to incorporate spatial
interactions. This allows us to verify that the power law
distribution of activities that is predicted by the non-spatial
model is retained when space is introduced. Furthermore, it allows
us to better map model output to empirical measurements.
Scale-free spatial network models that have been studied recently
are not directly applicable to urban economics since they require
an \emph{a priori} distribution of node
locations\cite{Barthelemy_2002,Brunet_Sokolov_2002,Manna_Sen_2002}.

\subsubsection{\label{tm_tsm_evolution}Network evolution}

In the continuum non-spatial model we were able to treat network
evolution as a purely local phenomenon since interactions in the
network are homogenous, see Eq. (\ref{eq:meanBA}). In the spatial
model we have to again explicitly separate the dynamics of the
edge end-points. This is because the spatial context of the first
end-point of a new edge modifies the probabilities with which
other sites will become the second end-point (transportation
costs).

The same types of growth that are used in the non-spatial model
are also used in the spatial model. But, as stated, the two
end-points here need to be dealt with separately. The first site
(primary increase) is selected either uniformly or preferentially
depending on growth type. This will be discussed shortly. The
second end-points of the trade relation edges (secondary increase)
are always preferential.

\subsubsection{Secondary increase}
The spatial model tells us how a given primary increase $p_j$ in
activity at site $j$ causes secondary increases $s_i$ in activity
at all other sites $i$ where
\begin{equation}
s_i=p_j\frac{D_{ij}x_i}{\sum_kD_{kj}x_k}, \label{eq:interact}
\end{equation}
This is analogous to Eq. (\ref{eq:BA}) but in the spatial model,
transportation costs will bias the choice of trade partners.
$D_{ij}$ is a matrix representing the interaction strength, which
is assumed to decay with the transportation costs. For the results
in this paper we have used $D_{ij}=(1+c d(i,j))^{-\alpha}$, where
$d(i,j)$ is the Euclidean distance between sites $i$ and $j$. The
non-negative parameters $c$ and $\alpha$ controls the impact of
spatial. Another function that can apply is
$D_{ij}=max(0,1-d(i,j))$ if the transportation characteristics of
the activity is known, which can be the case in a model where
activity types are modelled separately rather than as an average.
Exponential $D_{ij}=e^{- d(i,j)}$ can also be interesting to the
extent that shielding is important, i.e. an activity tends to
trade exclusively with the nearest supplier.

\subsubsection{Primary increase}
When trade takes place there is a mutual benefit in efficiency
that is often used for further increasing the activity in the
city. This feedback process makes it possible for the amount of
activity in the urban system to increase considerably faster than
the population. It is useful to think of Eq. (\ref{eq:interact})
as a black box system to which a driving force, primary increase,
is applied. It should also be noted that the primary effects used
for the model we present here are by no means neither exhaustive
nor final: most earlier models of urban growth could be introduced
as primary effects in our framework.

\subsubsection{Primary effects in type-1, type-2 and type-3 growth}

Network evolution in the spatial model is similar to the
non-spatial model. Secondary increases are always preferential
following Eq. (\ref{eq:interact}) and primary increases now
reflect the spatial distribution:

Type-1 growth: the primary uniform increase is identical to the
non-spatial case following Eq. (\ref{eq:UNI}).

Type-2 growth: the primary preferential increase is identical to
the non-spatial case following Eq. (\ref{eq:BA}).

Type-3 growth is split into two related processes where one
corresponds to growth in the perimeter of clusters and the other
corresponds to growth in connection to infrastructure in the rural
areas between clusters. Such infrastructure is not explicitly
represented in our model and instead we use a parameter $\epsilon$
to tune the amount of ambient infrastructure and thus the rate
with which seemingly isolated clusters will appear.

Type-3a growth: with a probability of $q_3 (1-\epsilon)$ we set
the activity of a perimeter node to $m$. Perimeter nodes are nodes
that are not developed but borders to at least one developed cell.
The site of the new node is selected randomly and with uniform
probability among the perimeter sites on the grid
\begin{equation}
\pi^a_i=\delta^{(P)}_i \frac{1}{n^{(P)}_t}  \label{eq:UNI_edge},
\end{equation}
where $\pi^a_i$ is the probability with which node $i$ is selected
to undergo type-3a growth, $\delta^{(P)}_i=1$ if the node $i$ is
on the perimeter, $\delta^{(P)}_i=0$ if the node is not on the
perimeter and $n^{(P)}_t$ is the number of perimeter nodes at time
$t$.

Type-3b growth: With a probability of $q_3 \epsilon$ we set the
activity of an external node to $m$. An external node is a node
that is undeveloped and that has no developed neighbors. The site
of the new node is selected randomly and with uniform probability
among external sites on the grid
\begin{equation}
\pi^b_i=\delta^{(E)}_i \frac{1}{n^{(E)}_t}
\label{eq:UNI_external},
\end{equation}
where $\pi^b_i$ is the probability with which node $i$ is selected
to undergo type-3b growth, $\delta^{(E)}_i=1$ if the node $i$ is
external, $\delta^{(E)}_i=0$ if the node is not external and
$n^{(E)}_t$ is the number of external nodes at time $t$.

\subsubsection{\label{tm_continuumspat}A continuum formulation of the spatial model}
A continuum formulation for the evolution of developed nodes in
the spatial model can, as in section \ref{tm_formulation_2}, be
constructed by studying the time evolution of expected node
degrees.
\begin{equation}
x_i(t+1)=x_i(t)+E[p_i(t)]+E[s_i(t)],
\end{equation}
where $E[s_i]$ is the expected secondary increase which can be
calculated by a weighted summation over the expected primary
increments,
\begin{equation}
\label{eq:expectsec}
E[s_i]=\sum_jE[p_j]\frac{D_{ij}x_i}{\sum_kD_{kj}x_k}.
\end{equation}
For the special case of our simple model for the primary effects
we have
\begin{equation}
E[p_j]=\delta^{(D)}_j(p_j^{(1)}+p_j^{(2)})+\delta^{(P)}_jp_j^{(3a)}+\delta^{(E)}_jp_j^{(3b)},
\end{equation}
with
\begin{eqnarray}
p_j^{(1)}=q_1\frac{m}{n^{(D)}_t}, \\
p_j^{(2)}=q_2\frac{mx_j}{\sum_kx_k}, \\
p_j^{(3a)}=q_3(1-\epsilon)\frac{m}{n^{(P)}_t}, \\
p_j^{(3b)}=q_3\epsilon\frac{m}{n^{(E)}_t}.
\end{eqnarray}
The total growth for a developed node can in this case be
separated into a uniform and a preferential part as
\begin{equation}
x_i(t+1)=x_i(t)+\zeta_i(t)+\eta_i(t)x_i(t), \label{eq:totgrowth}
\end{equation}
with
\begin{equation}
\zeta_i=q_1\frac{m}{n_t^{(D)}},
\end{equation}
and
\begin{equation}
\eta_i=\frac{q_2m}{\sum_kx_k}+\sum_jE[p_j]\frac{D_{ij}}{\sum_kD_{kj}x_k}.
\end{equation}
Noting that $\sum_i\eta_ix_i=m(q_1+2q_2+q_3)$, equation
\ref{eq:totgrowth} can be rewritten as
\begin{equation}
x_i(t+1)=x_i(t)+\zeta_i(t)+m(q_1+2q_2+q_3)\frac{\eta_i(t)x_i(t)}{\sum_j\eta_j(t)x_j(t)},
\end{equation}
which, in a comparison with equation (\ref{eq:meanBA}), reveals
that the only difference between the the spatial and the
non-spatial model is the site and time-dependent parameter
$\eta_i$. This is similar to the concept of node fitness, as
presented in \cite{Bianconi_Barabasi_2001a,Ergun_Rodgers_2001},
which can affect the node degree distribution. However, our
simulation results indicate that $\eta_i$ falls within a
sufficiently narrow interval for the power law to be essentially
preserved (Fig. \ref{price_cell_histogram}). This is also
supported by calculations of $\eta_i$ for both simulated and
empirical data.

\subsection{\label{tm_uc}The network model in an urban economics context}
\subsubsection{\label{tm_formulation_3}The connection between node degree and land value}

An approximate linear relationship between node degree in the
model and land value in the real system is crucial for the
interpretation of our results. The motivation follows from i)
market pricing of goods and services and ii) the connection
between trade benefits and land value.

 i) Market pricing of commodities provides an adaptive measure
that allows us to compare the activities that generate them.
Hence, on average, an edge contributes identically to the value of
both nodes to which it connects. This contribution is exactly our
definition of activity, which implies that the degree of a node is
proportional to its benefits due to trade.

ii) This connection consists of two proportionalities. For a node
$i$ we have
\begin{equation}
v_i \propto r_i \propto x_i,
\end{equation}
where $v_i$ is the value of the corresponding land area, $r_i$ is
the bid-rent\cite{Alonso_1960,Alonso_1964}, and $x_i$ is the total
trade benefits as outlined above. Capitalizing periodic rent
income from the site $i$ gives land value $v_i=\frac{r_i}{i}$
where $i$ is the interest rate\cite{OSullivan_2002}. The second
proportionality is a weak form of the leftover principle from
urban economics, which states that, in a competitive land market,
rent equals the amount of money leftover after all expenses
(except rent) are paid. This amount of money equals the sum of all
trade benefits at the site. For our results it is sufficient that,
on average, a certain proportion of each new unit of trade benefit
goes to the landowner.

Together, i) and ii) suggest an approximately linear relationship
between node degrees and land values.

\subsubsection{\label{tm_formulation_4}Types of growth}
Urban activity can increase in essentially two ways: either a new
activity is related to, or it is unrelated to, an existing
activity at the site. In the former case (preferential growth)
this could be a new employee hired as a response to increased
demand, in the latter case (uniform) it could be the establishment
of a new firm. Preferential growth corresponds to a per-unit
activity rate. Uniform growth corresponds to establishment among
lots on a competitive land market where, for the average land use,
we can not expect any lot to be more profitable than any other.

\subsubsection{\label{tm_formulation_5}Reasons for treating perimeter growth separately from internal growth}

The jagged perimeter of urban areas exposes large areas of
undeveloped land to urban infrastructure, thus making it
attractive for urban land use. Because perimeter land is in ample
supply and currently has a low revenue, even land uses with a very
low trade gain can be competitive. Many low-activity land uses in
the outskirts of the urban area can likely just barely out-bid
agriculture and would not be viable in competition with other
urban land uses. Among high-activity land uses some have very
specific demands on land improvements and can therefore not
benefit from buying existing buildings inside the urban area. This
creates a special case for perimeter land. Note that just like for
type-1 growth, competition prevents prediction of where the next
growth event will take place among the perimeter nodes, and no
preferential growth is possible since there is no previous
activity that can expand.

\subsubsection{\label{sat_f}Node fitness and growth}

If we only regard trading activities the only difference between
two sites with identical activity is the value of their spatial
context. Therefore, in the spatial model, secondary growth is not
homogenous, see Eq. (\ref{eq:interact}). Node fitness (Eq.
\ref{eq:totgrowth}) can be viewed as a local interest rate that
predicts the growth rate of activity investments made in that
site. The result of this is non-trivial growth predictions since
the expected amount of new local development does not become a
simple fraction of current development.

Most notably, the model predicts the emergence of urban
sub-centers. This is realized by examining the expected secondary
effects in the spatial model (Eq. \ref{eq:expectsec}). Apart from
being proportional to the amount of present activity $E[s_i]$ is
also subject to site competition and will increase for nodes that
have high activity in relation to their own neighborhoods. For
each possible primary effect in $j$, the node development $x_i$ is
weighted with the fraction between trade intensity between site
$j$ and $i$ and the sum of the all other trading options for the
primary effect under consideration. Thus, nearby high-intensity
nodes will not necessarily benefit a small neighbor.

\section{\label{er}Results}
\subsection{\label{er_0}Land values are power law distributed}
From empirical data, land values in Sweden are demonstrated to
follow a power-law distribution for the higher range of land
values (see figure \ref{price_cell_histogram}). The sharp
transition that can be observed around $60$kSEK suggests that two
truly different mechanisms generate the prices below and above
this point. This is in agreement with the observation that the
pricing mechanism we suggest would apply only to trading activity.

The data we have used is based on the land value component of
market value estimations of about 2.9 million units of real estate
in Sweden. The data was originally compiled by the Swedish
National Land Survey and coded by Sweden Statistics to
geographical coordinates. The data points we use are aggregated
land values into $100m\times100m$ squares.

Our results are supported by a recent study by Kaizoji that shows
scale-free behavior of land prices in Japan, with an exponent
$\gamma$ ranging from $2.53$ to $2.76$\cite{Kaizoji_2003}.

\subsection{\label{er_1}Power law distributed prices are predicted by the non-spatial model}

We have developed a simple network model of the urban economy
based on the Barab\'asi-Albert model by mapping fundamental
assumptions from urban economics to the ontology of the network
model (see section \ref{tm_formulation}). In our derivation, we
have determined the exponent of the node degree distribution, and
thus, per our definition, also the predicted land values, to
follow equation (\ref{eq:exponent}). To the extent that our
interpretation of the underlaying dynamics is correct this
demonstrates why urban land values can be expected to follow a
power law and how the exponent may depend on parameters.

\subsection{\label{er_2}The spatial model retains power law statistics}

As discussed in section \ref{tm_continuumspat}, the impact of
spatial constraints closely resembles that of node
fitness\cite{Bianconi_Barabasi_2001a,Ergun_Rodgers_2001}. This
could potentially result in the distribution for the spatial
network model becoming a sum of many power laws with different
exponents.

\begin{figure}
\includegraphics[width=8.5cm]{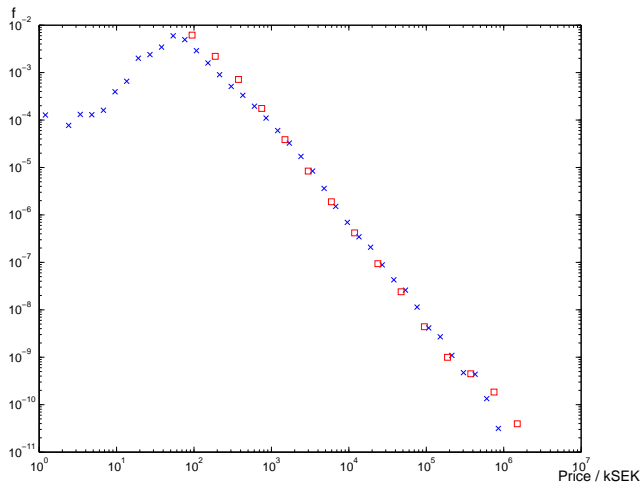}
\caption{\label{price_cell_histogram} \em Double logarithmic
histogram of simulation results and empirical data. Simulation
results are are denoted with squares in the figure and they are a
mean of the results of three runs of the spatial model. The
exponent of the model output has been tuned to match the empirical
data. As indicated in Eq. (\ref{eq:exponent}) this is done by
setting the relative proportions of the growth types, in this case
$q_1=0.1$, $q_2=0.6$, $\epsilon=0.01$, $c=0.2$, $\alpha=1$,
$m=100$kSEK and $t$=170000. The exponent is roughly $2.1$ which is
close to the value of $\gamma=2.25$ that is predicted by
Eq.(\ref{eq:exponent}) for these parameter values in the
non-spatial model. A slightly different value for the spatial
model must be expected because of spatial bias in the growth
dynamics. The sharp transition that occurs around a land price of
60kSEK ($\approx$ 5kUSD) per $100m\times100m$ marks the difference
in dynamics between trade based urban activities and rural
activities whose values are not described by the presented model.}
\end{figure}

In figure \ref{price_cell_histogram} we plot results from
simulations showing that node degrees in the spatial model follow
a power law distribution. The model parameters have been tuned
(see sec. \ref{tm_continuumspat}) to reproduce the distribution of
the observed land prices.

\begin{figure}
\includegraphics[width=8.5cm, height=4.0cm]{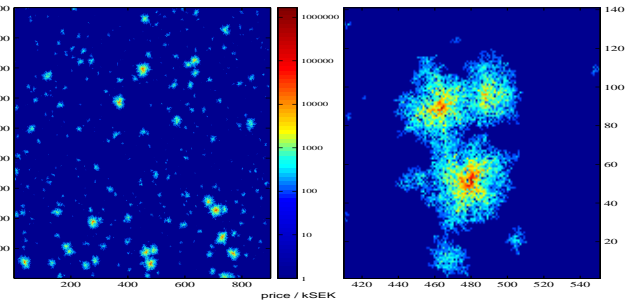}
\caption{\label{simulated_configuration} \em Simulated
configuration from the spatial model with the parameter values
used in previous plots.}
\end{figure}

\section{\label{conc}Conclusion}
We present empirical evidence that land prices follow a power law
distribution for urban land uses, and present a generic model,
based on the underlaying trade network, that reproduces this
behavior. The model is a version of the Barab\'asi-Albert scale
free network that is also extended to incorporate spatial
constraints.
%

The applicability of the network paradigm to urban growth suggests
that scale invariance in urban systems is caused by: i) growth and
ii) preferentiality in how new trade connections are formed
between areas of land. Growth in this context refers to the
continual development of new land. Preferentiality is a
consequence of point-to-point interactions between activities
occupying the land areas. Note that many other observables, such
as population and urban land use intensity, might be highly
correlated with land value.

The spatial model that we present can have more general
applicability beyond urban economics. Other spatially growing
networks are communication networks, transportation networks,
electricity and utility networks. These can be expected to follow
a similar type of growth since they are intimately connected to
urban activity.

The network architecture is generic and allows for addition of any
amount of detail. Also, being based on trade relations, the model
produces output in units of currency. Because of this, such
network models can provide a bridge between a microscopic dynamics
that can be found empirically and emergent economic properties.

Further possible directions for research on urban economic
networks include interpretation of other theoretical network
results in terms of urban dynamics, finding empirical parameters
for scenario predictions and model validation.

\section{Acknowledgements}
We would like to thank Martin Nilsson and Steen Rasmussen for
valuable input and discussions.

\vspace{2cm}

\bibliographystyle{unsrt}
\bibliography{../references}

\end{document}